\documentclass[doublecol]{epl2}
\title{Effective temperatures from the fluctuation-dissipation measurements in soft glassy materials }

\author{S. Jabbari-Farouji\inst{1,2,*}, D. Mizuno \inst{3,4,5,*}, D. Derks \inst{6}, G.H.\ Wegdam \inst{1}, F.C.\ MacKintosh \inst{2},  C.F.\ Schmidt \inst{3,5} and D.\ Bonn \inst{1,6}}

\institute{
  \inst{1} Van der Waals Zeeman Institut, Universiteit van
Amsterdam, Valckenierstraat 65, 1018 XE,  Amsterdam, The
Netherlands\\

 \inst{2} Theoretical Physics and Polymer Group, Department of Applied Physics, Technische Universiteit Eindhoven
 5600MB Eindhoven, The Netherlands\\

  \inst{3}Department of Physics and Astronomy, Vrije
Universiteit Amsterdam, De Boelelaan 1081, 1081HV Amsterdam, The Netherlands\\

\inst{4}Organization for the Promotion of Advanced Research, Kyushu University, Higashi-ku, Hakozaki 6-10-1, 812-0054 Fukuoka,
Japan\\

\inst{5} Physikalisches Institut, Georg-August-Universit\"{a}t, Friedrich-Hund-Platz 1, 37077 G\"{o}ttingen, Germany\\

 \inst{6} Laboratoire de Physique Statistique, Ecole Normale
Sup\'erieure, 24, rue Lhomond, 75231 Paris Cedex 05, France\\

\inst{*} Both authors have equally contributed to this work.}

\pacs{nn.mm.xx}{First pacs description}
\pacs{nn.mm.xx}{Second pacs description}
\pacs{nn.mm.xx}{Third pacs description}

\abstract{We have investigated the validity of the
fluctuation-dissipation theorem (FDT) and the applicability of the concept of effective temperature in a number of
non-equilibrium soft glassy materials. Using a combination of
passive and active microrheology to measure displacement
fluctuations and the mechanical response function of probe
particles embedded in the materials, we have directly tested the validity of the
FDT. Our results show no violation of the FDT over several decades in frequency (1-10$^4$ Hz) for hard sphere colloidal glasses and colloidal glasses and gels of Laponite. We further extended the bandwidth of our measurements to lower frequencies (down to 0.1 Hz) using video microscopy to measure the
displacement fluctuations, again without finding any deviations from the
FDT.}

\begin{document}

\maketitle

\section{Introduction}

While the foundations of equilibrium statistical mechanics are
well-established, non-equilibrium statistical physics
is still being developed. One attempt in the direction of
developing a statistical mechanical description of
non-equilibrium, but slowly evolving systems is to extend the fluctuation dissipation theorem (FDT) to non-equilibrium situations \cite{cugliandolo}.
The FDT relates the relaxation of spontaneous fluctuations to the response of an equilibrium system to a weak external
perturbation. Specifically, the FDT states that the response function
is proportional to the power spectral density of the thermal
fluctuations, with a prefactor given by the temperature. To describe non-equilibrium systems, the idea is to relate
the (non-equilibrium) fluctuations to the response function via
a (possibly time- or frequency-dependent) effective temperature $T_{\mbox{\scriptsize eff}}$ \cite{cugliandolo}.
 Violations of the equilibrium FDT have recently been
studied extensively in theory and simulations for various model systems such as structural glasses \cite{parisi,kob1}, spin glasses \cite{cugliandolo,spin1,trapfdt,theo1} and driven systems such as a
fluid under shear \cite{driven} and aging critical systems
\cite{agingcritical}.

The experimental situation, however, is less conclusive.
Experiments have been reported on structural glasses \cite{Israel}, spin glasses
\cite{spinexp}, granular matter \cite{granular}, hard sphere
colloidal glasses \cite{HSdaniel,HSMakse} and the colloidal glass
of Laponite \cite{Ciliberto,fdtabou,FDTPRL}, with sometimes contradictory
results.

For the 'soft' systems considered here, the experimental situation is particularly confusing. For hard sphere colloidal fluids there have been two reports on violation of FDT with very different results. Bonn et. al. determined a $T_{\mbox{\scriptsize eff}}$ that almost reached $100$ and depends strongly on frequency \cite{HSdaniel} whereas a Wang et. al report a constant effective temperature of twice the bath temperature \cite{HSMakse}.

Laponite is another glassy system for which contradictory results
have been reported.  Bellon et.\ al.\ \cite{Ciliberto}  reported
an effective temperature  from electrical measurements indicate a
strong violation of the FDT in the frequency range 1-40 Hz,
whereas the same group did not see deviations for the mechanical
measurements in the frequency range 1-20 Hz \cite{Ciliberto}. Abou
et.\ al.\ also reported an effective temperature different from
the bath temperature \cite{fdtabou}. In contrast to these
findings, our simultaneous measurements of displacement
fluctuations and mobility using microrheology in the same system
show no deviations from the FDT over several decades in frequency
(1-10$^4$ Hz) and for all aging times: $T_{\mbox{\scriptsize
eff}}=T_{\mbox{\scriptsize bath}}$ \cite{FDTPRL}. Therefore
experimentally the usefulness of the extension of the FDT to
non-equilibrium situations and the concept of effective
temperature is still a matter of controversy and deserves further
investigation. One of the main difficulties seem to be that the
direct and simultaneous measurement of the fluctuation correlation
function and the
 and corresponding response function is experimentally rather cumbersome and requires elaborate and high-resolution experimental techniques.

To test the FDT directly, one can use microrheology to measure
both the position fluctuations and the complex compliance of a
probe particle simultaneously. There are two classes of
microrheology (MR) technique: \emph{active} and \emph{passive}. In
passive MR, one measures the displacement fluctuations of a probe
particle: $x(t)$. Then, assuming that the FDT holds, one can
obtain the imaginary part of the response function
$\alpha''(\omega)$ from the power spectrum of thermal fluctuations
$\langle|x(\omega)|^2\rangle$ \cite{Mason,Gittes}. Passive MR was
first introduced by Mason et. al. \cite{Mason,MasonDWS97} and was
already then used in non-equilibrium system of hard sphere
glasses. Their conclusion was that the complex shear modulus obtained from the passive MR
 qualitatively to be in agreement with the bulk rheology.
 In the active method \cite{FDTPRL,Yang,Mizuno}, one directly measures $\alpha(\omega)$,
the mechanical response of the probe particle to an applied
oscillatory force. Consequently, by comparing the power spectrum
of thermal fluctuations $\langle|x(\omega)|^2\rangle$ with the
imaginary part of the response function $\alpha''(\omega)$, one
can obtain the frequency-dependent effective temperature over a
wide range of frequencies (0.1Hz-10$^4$ Hz) for soft glassy
materials from:
\begin{equation}
\label{eq:Teff}\frac{ T_{\mbox{\scriptsize
eff}}}{T_{\mbox{\scriptsize bath}}}=\frac{\omega
\langle|x(\omega)|^{2}\rangle}{2k_{B}T_{\mbox{\scriptsize bath}}
\alpha_{\mbox{\scriptsize
active}}''(\omega)}
\end{equation}
If $T_{\mbox{\scriptsize eff}}=T_{\mbox{\scriptsize bath}}$, then
the fluctuation-dissipation relation is satisfied. The combined use of active and passive MR thus provides a way to directly test the applicability of the FDT in non-equilibrium systems.

In this Letter, in view of the existing controversies in the literature, we test the validity of the FDT using microrheology  on hard sphere colloidal glasses and aging colloidal gels and glasses of Laponite. We do not find any deviations from FDT in any of these systems over the entire range of frequencies probed in the experiments.

\section{Experimental}

Laponite XLG suspensions in pure water were prepared as explained
in \cite{FDTPRL}. Samples containing salt were prepared by
diluting the Laponite suspensions in pure water and adding a
concentrated salt stock solution \cite{Nicolai2000}. Immediately
after the preparation of the samples, we mixed in a small
fraction, below $ 10^{-4}$ vol \%, of silica probe beads. This
instant defines time zero of the "waiting time". After mixing we
infuse the suspension into a sample chamber of about 50 $\mu$l
volume, consisting of a coverslip and a microscope slide separated
by spacers made of UV glue with a thickness of about 70 $\mu$m and
sealed with vacuum grease at the ends to avoid evaporation of
sample.Concentrated hard-sphere-like suspensions were prepared by
dispersing sterically stabilized PMMA colloids of a diameter of
397 nm and a polydispersity of 10\% \cite{didi} in a nearly
refractive-index and density-matching mixture of cyclohexyl
bromide and \textit{cis-}decalin (w/w = 3:1). The values for the
density and the refractive index of the solvent mixture are nearly
the same as those reported for PMMA ($\rho$ = 1.19 g/cm$^3$ and
the index $n = 1.49$ \cite{didi}). To screen the charges on the
colloidal particles, the mixture was saturated with tetrabutyl
ammonium bromide (concentration 300 $\mu$M). The shear viscosity
of the solvent mixture was measured with a rheometer as
$\eta_s=2.47$ mPa.s. As probe particles we used 1.1 $\mu$m in
diameter melamine resin particles (micro-particles GmbH, Germany)
with a density $\rho = 1.51$ g/cm$^3$ and refractive index $n =
1.68$; the high index makes it possible to optically trap them. We
prepared the samples by weight fraction, assuming this to be equal
to the volume fraction. The suspension was subsequently introduced
into a sample chamber and sealed with epoxy glue to avoid
evaporation of the sample. After placing the sample chamber in the
microscope, we trapped a single bead, moved it to 20-30 $~\mu$m
above the glass surface, and then performed the active and passive
MR experiments with it. All experiments were performed at room
temperature $(21\pm 1^\circ $C).

The microrheology setup is  described in detail elsewhere
\cite{Mizuno, Maryam}. In brief, it consists of two optical
tweezers formed by two independent polarized laser beams that are
superimposed. One of the lasers is used to drive the oscillations
of the trapped particle (drive laser, $\lambda=$1064 nm. The other
one is stationary and is used to detect the position of the
particle (probe laser, $\lambda=830$ nm). Stable trapping is
achieved using a high numerical aperture objective lens. The
interference signals of the laser beams emerging from the
condenser lens (1.4 NA) after passing through the sample are
projected onto quadrant photodiodes (QPD).

In the active measurements, the driving laser trap was oscillated
through an acousto-optical deflector (AOD) and the output signal
from the QPD that detected the probe laser was fed into a lock-in
amplifier. Thus we were able to measure the amplitude and phase of signals resulting from driving laser displacements.

We thus measure phase and
displacement amplitude of the particle in response to the
oscillatory force exerted in the $y$ direction. We oscillated the
drive laser with frequency $f=\frac{\omega}{2\pi}$ and an
oscillation amplitude $L$ ($\approx 70$ nm)  . This displacement results in an
oscillatory force of magnitude $\Re{\{k_{d}[L \exp(-\imath \omega
t)-x(t)]\}}$ on the bead, where $k_{d}$ refers to the trap
stiffness of driving laser.  Here, $f_{0}=k_{d} L \exp(-\imath \omega
t) $ can be interpreted as the apparent oscillatory force exerted
by the drive laser. The particle displacement caused by the
apparent oscillatory force is denoted as $x_{\omega}(t)$. The
apparent response to the sinusoidal force is defined as
\begin{equation}\label{eq:alpha}
 <x_{\omega}(t)>= \alpha_{\mbox{\scriptsize app}}(\omega) f_{0}
\exp(-\imath \omega t)
   \end{equation}
In practice the lock-in amplifier measures the particle
displacements at the oscillation frequency, $<x_{\omega}(t)>=
\tilde x\exp(-\imath \omega t)$. It can
be shown that the apparent complex compliance, which includes the influence of the trap, is \cite{Mizuno}
\begin{equation}
\label{eq:a12}
 \alpha_{\mbox{\scriptsize app}}=\frac{\tilde x}{f_{0}}=\frac{1}{k_{d}-\imath \omega
 \xi}
   \end{equation}
where $\xi$ is a friction coefficient reflecting the memory
effects in viscoelastic materials. The true response of the embedding material is $\frac{1}{-\imath \omega \xi}$.
Therefore the apparent complex compliance must be corrected as
$\alpha=\frac{\alpha_{\mbox{\scriptsize app}}}{1-k_d \alpha_{\mbox{\scriptsize app}}}$.

Analyzing the data in the active method requires knowing the trap stiffness and
calibration of the measured amplitude of response to absolute distances. It also requires a
correction of the measured amplitude and phase of the response given the response characteristics of the AOD  the details of which can be found in \cite{Mizuno}.
The trap stiffness and the calibration factor for converting voltage to nanometers for both driving laser and
probe laser can be found from the spectra of beads in pure solvents \cite{Maryam}.

Since the drive laser was relatively strong ($\approx 50-100$ mW depending on the experiments), there was the possibility that heating effects influence the experiment. Local heating of the solution can lead to a refractive index gradient that can
act as an optical lens causing the deflection of the probe laser;
the so called photo-thermal lens effect. We tested for such an effect by
measuring the AOD response in an our samples in the absence of
a bead and found the effect to be negligible (data not shown).

\section{Results}
\begin{figure}[t]
\includegraphics[scale=0.75]{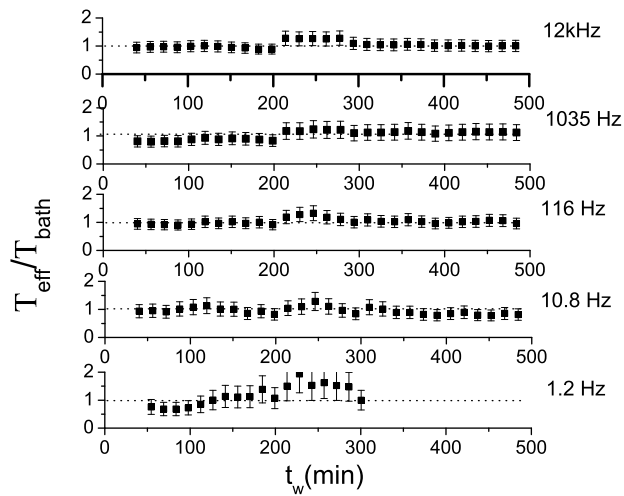}
\caption{ Effective temperature in Laponite
 glass (2.8 wt\% concentration). Ratio of (apparent) response derived from
fluctuations via FDT and actively measured response
${T_{\mbox{\scriptsize eff}}}/{T_{\mbox{\scriptsize bath}}}$ as a
function of waiting time for frequencies $f=12000,
1035,116,10.8$ and 1.2 Hz (from top to bottom). Note that the apparent
increase after 200 min, is due to calibration errors, caused by
changing the trap power to apply the oscillatory force.
  }\label{fig1}
\end{figure}

We have first studied the validity of the FDT in an aging colloidal glass of Laponite, as reported in detail in \cite{FDTPRL}. We do this by comparing the response functions obtained from active (direct) and passive (indirect) microrheology, where both active and passive measurements are performed on the same bead. The results are plotted in terms of an effective temperature from Eq.\ (\ref{eq:Teff}) in Fig.\ \ref{fig1}. The figure shows that the effective temperature was equal to the bath temperature to within the experimental uncertainty over a wide range of frequencies $f\approx1-10^4$ Hz for all the
measured waiting times.

Colloidal gels are another class of non-equilibrium systems showing aging phenomenon. In
such systems, similarly to what happens in glasses, correlation
and response functions can be a function of time elapsed since
their preparation and their relaxation times tend to grow in time. Laponite colloids, within a certain range of colloid concentrations and salt content, can form soft colloidal gels which evolve from an
initially liquid-like state to a viscoelastic solid-like state
\cite{PRL,newpaper}. The aging mechanism behind the gel formation is different from that of a glass leading to a  spatially heterogenous structure in colloidal gels \cite{PRL,longMR,newpaper}.

In  a recent work \cite{longMR} we have shown that the heterogenous structure of colloidal gels can be detected by passive MR measurements.
Displacement PSDs  at intermediate stages of aging measured at almost the same time but different positions of a gel
were not equal, hence demonstrating the heterogeneity of the samples at length scales of a few
micrometers, contrary to what was found in the glassy samples. Furthermore, at some positions in the gel samples we
observed some anisotropy, i.e.\, displacement PSDs measured at $x$
and $y$ directions were not equal.

Despite a number of studies on the validity of FDT
in colloidal glasses, there has, as far as we know, as yet been no
study investigating the FDT in aging non-equilibrium gels. Here we would like to test the validity of FDT locally at different positions of a gel in spite of the fact that measured PSDs can vary from one point to the other point.

We tested the validity of the FDT in an aging sample of 0.8 wt \% Laponite with 3 mM added salt NaCl
which is known to be in the gel range \cite{newpaper}.
To this end, we performed two sets of experiments. First we investigated the
FDT at a fixed position, using one and the same probe bead, in the sample at different
stages of aging. Similar to the aging colloidal glass, the actively
and passively measured responses were in agreement with each other up to the maximum
waiting times of 100 h. In a further set of experiments we chose a few beads
at different positions in the sample and investigated the validity
of the FDT at late stages of aging by performing
both active and passive microrheology on each bead in the sample.

\begin{figure}[h!]
\begin{center}
\includegraphics[scale=0.4]{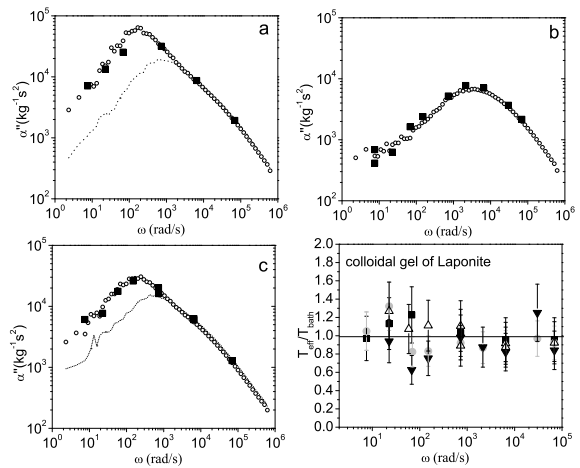}
\caption{ a,b and c) Imaginary parts $\alpha''$ of the frequency-dependent
response functions in the $y$ direction from active (filled
symbols) and passive (open symbols) microrheology using 0.5 $\mu$m
diameter silica beads at 3 different positions of the 0.8
wt\% Laponite (3mM NaCl) sample measured 5 days after sample preparation. The dotted lines in panel (a) and (c) show
the $\alpha''$ in the $x$ direction which is different from its
value in the $y$ direction at these positions of the sample. d) Ratio of effective temperature to bath temperature from passive and active MR
${T_{\mbox{\scriptsize eff}}}/{T_{\mbox{\scriptsize bath}}}$ as a
function of frequency, obtained with 0.5 $\mu$m diameter silica
beads at 4 different positions (shown by the different symbols) in the sample.
 }\label{fig4}
\end{center}
\end{figure}

Fig.\ \ref{fig4} shows  the response functions obtained from active
and passive microrheology measured at 3 different positions in 0.8 wt\%
Laponite (with 3 mM NaCl) 5 days after sample preparation.  The rate of aging for this sample was, at that time, so slow that there was no significant evolution during one measurement. For beads (a) and (c) the measured PSDs are not equal in the $x$ and $y$ directions. Nevertheless the apparent response from passive
microrheology in the $y$ direction agreed well with active
response obtained by exerting an oscillatory force in the $y$
direction.

The results show that, despite the presence of
heterogeneity in the gels, the FDT in the form of the Einstein
relation is valid locally at each point of the sample just as it
was for the glassy system.

Next we tested the validity of the FDT in a hard sphere colloidal glass ($\Phi\geq 0.58$) and a number of supercooled fluids ($ 0.49 <\Phi < 0.58$) as there have also been reports of FDT violation for such
systems \cite{HSdaniel,HSMakse}. Note that supercooled liquids are in principle ergodic systems and one does therefore not expect to see
deviations for the FDT in these systems. However, to resolve the
controversy existing in the literature \cite{HSdaniel}, we have done experiments also on such ``supercooled" liquids. First, we
found that the PSDs of beads measured at different positions in the samples were equal, i.e. that the samples were spatially homogeneous on the micrometer scale. In the measured frequency range, we furthermore
did not see any effect of aging (no change of PSDs) in the colloidal glass within the 12h duration of the experiments.
\begin{figure}[h!]
\begin{center}
\includegraphics[scale=1]{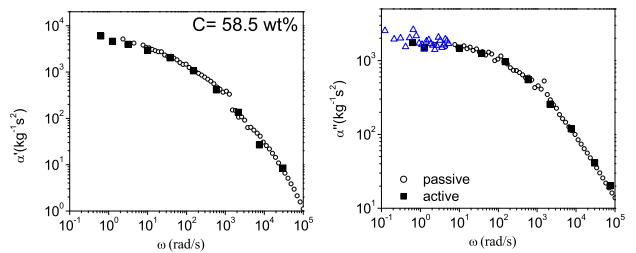}
\caption{ Real  $\alpha'$ and imaginary parts $\alpha''$ of the
frequency-dependent response function from active (filled symbols)
and passive (open symbols) microrheology using 1.1 $\mu$m diameter
melamine probes for a hard sphere colloidal glass with volume fraction
$\phi=0.585$. Data at low
frequencies $\omega <10$ rad/s (triangles) were obtained from video microscopy. }\label{fig2}
\end{center}
\end{figure}

\begin{figure}[h!]
\begin{center}
\includegraphics[scale=0.6]{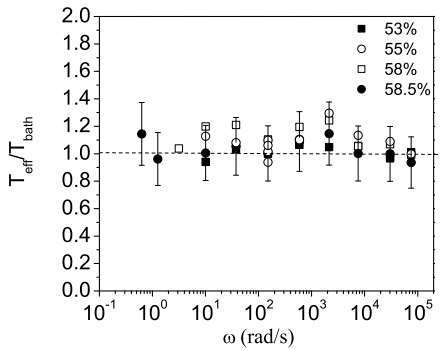}
\caption{The effective temperature as a function of frequency for
53 (supercooled fluid), 58 and 58.5 wt \%  (glass) $d=397$nm colloids and 55 wt \% (supercooled fluid)
concentration  of $d=1.13 \mu$m colloids . The concentrations are
shown in the legend. The samples in the supercooled region
were measured within a couple of hours after preparation, since at
later times, we observed the formation of crystals. The glassy samples were measured both immediately after preparation and after 12h.
 }\label{fig3}
\end{center}
\end{figure}

As evident in Fig.\ \ref{fig2}, the agreement between
actively and passively measured response functions is excellent. This means again
that no deviations from the FDT were observed over the range of
frequencies the measurements covered. To extend the range of our measurements to even lower
frequencies we have used particle tracking in video microscopy to obtain the PSD for
$0.1 \leq \omega < 1$ Hz. As can be seen in Fig. \ref{fig2} there was
again a very good agreement between actively and passively measured response.
In Fig.~\ref{fig3}, we plot the resulting ratio ${T_{\mbox{\scriptsize
eff}}}/{T_{\mbox{\scriptsize bath}}}$, as defined in Eq.\ (\ref{eq:Teff}) as a function of frequency for the two glassy samples and two supercooled
liquids. This result confirms again that to within the experimental
uncertainty, the FDT is valid in the measured range of frequencies from 0.1 Hz to 10$^4$ Hz
and that the effective temperature does not differ from the bath
temperature. This is in agreement with the earlier experiments of Mason
for a glassy sample of hard spheres $\phi\approx 0.56$ who compared passive MR to bulk rheology over a frequency range $0.1 < \omega < 100$ Hz
\cite{MasonDWS97}.

\section{Discussion and Conclusions}

We have investigated the validity of the FDT in colloidal glasses and gels of Laponite and in a hard-sphere colloidal glass in the frequency range
$0.1-10^4$ Hz. We see in each case a good quantitative agreement between the actively measured
response function and the correlation function of spontaneous thermal fluctuations,
implying that there is no measurable violation of the FDT in any of these
non-equilibrium systems. Equivalently, we find in all cases an effective
temperature that does not differ from the bath temperature over the
measured frequency range.

Our results disagree with the
simulation results which observe deviations from FDT in a
 binary Lennard-Jones mixed glass \cite{kob1} and in a fragile glass
\cite{parisi}.  In both of these systems the deviations from FDT have
been observed at intermediate and long times, when
the decay of correlation functions is not exponential anymore. 
From the former studies \cite{SIF}, it is possible to estimate the frequency range where we should have expected the violation if that ever existed.
To do so, we rescale the frequency with the Brownian time scale
$\tau_B=\frac{R^2}{6D_0}$. For our hard sphere colloidal glass with particle
radius of 197 nm, $\tau_B=1.46\times 10^{-2}$ s. Therefore the
range of dimensionless frequencies in our measurements vary as
$8.7\times 10^{-3}<\omega \tau_B<1.46\times10^3$. In the data of van Megen
et al. \cite{SIF} for a colloidal glass similar to ours, the
plateau region of the intermediate scattering function was observed
for time scales $1<t/\tau_B<10^5$ which corresponds to $6\times
10^{-5}<\omega \tau_B<6$. Therefore at least part of the frequency
range of our measurements is in the plateau region corresponding to the time regime for
which the models suggest that the FDT violations should be
visible.  Besides the obvious difference between such 'real' glasses and soft glasses considered here, we have no explanation for this discrepancy.

The same experimental systems as those studied here had been
investigated by other groups, as we mentioned in the introduction.
The rheological measurements of the Ciliberto group in a colloidal
glass of Laponite are in agreement with our results whereas their
electrical measurements disagree \cite{Ciliberto}. The reason for the
disagreement could be that in the electrical measurements a different degree of
freedom was monitored. In contrast to this, simulations on a sheared binary
Lennard-Jones fluid \cite{driven} have shown that the effective
temperature are independent of the chosen observable in such systems. To confuse matters even more, it has
been shown theoretically \cite{trapfdt} that the effective
temperature in the glass phase of the Bouchaud's trap model does depend on the observable.
Subsequent work of Cilibetro et al on the Laponite \cite{Ciliberto} (2003) showed, however, that the observed violations of electrical FDT were due to violent and intermittent events with unknown origin.

Abou et al found violations of FDT during the aging of Laponite system by using an
experimetnal approach similar to ours \cite{fdtabou}.  Our method, however, is more direct, since
we are measuring the fluctuations and the response with the very same probe
particle. 

In hard sphere glasses there have also been reports of
violations of the FDT. The large violations reported by Bonn and Kegel \cite{HSdaniel} are likely to be due to the fact that they compared rheological measurements and light scattering experiments from two different systems, although both were in principle hard sphere systems. Furthermore, they did not use the correct form of Stokes formula in their analysis.
The Makse group \cite{HSMakse} has reported an
effective temperature for a hard sphere glass ($R=1.5 \mu$m) that
was twice the bath temperature at long times $t\sim 1000s$.
This time corresponds to a dimensionless frequency $\omega
\tau_{B}\approx 4\times 10^{-2}$, which is comparable to our
lowest measured frequencies $\omega \tau_{B}\approx 10^{-2}$. 
In \cite{HSMakse}, mobility was determined as the ratio of a constant force
to the response displacement of a particle, which is not generally accepted way to
estimate the effective temperature (especially for glassy materials taking infinite time to relax).

Recently Greinert et al. \cite{Bartlett} also have reported a $T_{\mbox{\scriptsize eff}}$ larger than bath temperature for an aging colloidal glass of Laponite at late stages of aging. Although their method using equipartition theorem rather than FDT might be useful to investigate the nonequilibrium signature at low frequencies, technically it is difficult to exactly rule
out the low fequency noise from the displacement fluctuation, which leads to apparent $T_{eff}$ higher than $T_{bath}$. In fact Jop. et al. have performed measurements using the same method as  Greinert et al. and find an effective temperature equal to bath temperature \cite{Jop}, suggesting that inferring an effective temperature necessitates a careful treatment of data at long waiting times.

In conclusion, we have here applied a technique that offers
a direct method for the simultaneous measurement of response and
correlation functions. This overcomes difficulties for experiments with non-equilibrium soft materials or glasses which have hampered earlier approaches searching for an effective temperature different from the ambient temperature.  Our experiments on glasses, gels and supercooled liquids  show no deviations from FDT.

\acknowledgments This research was supported by the Foundation for Fundamental Research on Matter
(FOM) of the Dutch  NWO.
LPS de l'ENS is UMR8550 of the CNRS , associated with university
Paris 6 and 7. Further support (C.F.S.) came from the DFG Center for
the Molecular Physiology of the Brain (CMPB) and the
Sonderforschungsbereich 755 of the German Science Foundation (DFG).


\end{document}